\documentclass[prb,twocolumn,showpacs,preprintnumbers,amsmath,amssymb]{revtex4}

\usepackage{graphicx}% Include figure files
\usepackage{dcolumn}% Align table columns on decimal point
\usepackage{bm}% bold math
\usepackage{epic}

\begin{document}

\preprint{APS/123-QED}

%\title{Magnetic properties of small Fe, Co and Ni clusters on Ir(111), Pt(111) and Au(111)}
\title{Trends in the magnetic properties 
       of Fe, Co and Ni clusters and monolayers on Ir(111), Pt(111) and Au(111)}

\author{S.~Bornemann$^1$, O.~\v{S}ipr$^2$, S.~Mankovsky$^1$, S.~Polesya$^1$, 
        J.B.~Staunton$^3$, W.~Wurth$^4$, H.~Ebert$^1$ and J.~Min\'ar$^1$}
\affiliation{$^1$Department Chemie und Biochemie, 
                 Ludwig-Maximilians-Universit\"at M\"unchen, 
                 81377 M\"unchen, 
                 Germany \\
             $^2$Institute of Physics of the ASCR v.~v.~i., 
                 Cukrovarnick\'{a}~10, 
                 CZ-162~53~Prague, 
                 Czech Republic\\
             $^3$Department of Physics, 
                 University of Warwick, 
                 Coventry CV4 7AL, 
                 United Kingdom\\
             $^4$Institut f\"ur Experimentalphysik, 
                 Universit\"at Hamburg and 
                 Centre for Free-Electron Laser Science, 
                 22761 Hamburg, 
                 Germany}
\email{sven.bornemann@cup.uni-muenchen.de}

\date{\today}

\begin{abstract}

We present a detailed theoretical investigation on the magnetic properties of
small single-layered Fe, Co and Ni clusters deposited on Ir(111), Pt(111) and
Au(111). For this a fully relativistic {\em ab-initio} scheme based on density
functional theory has been used.  We analyse the element, size and geometry
specific variations of the atomic magnetic moments and their mutual exchange
interactions as well as the magnetic anisotropy energy in these systems.  Our
results show that the atomic spin magnetic moments in the Fe and Co clusters
decrease almost linearly with coordination on all three substrates, while the
corresponding orbital magnetic moments appear to be much more sensitive to the
local atomic environment. The isotropic exchange interaction among the cluster
atoms is always very strong for Fe and Co exceeding the values for bulk bcc Fe
and hcp Co, whereas the anisotropic Dzyaloshinski-Moriya interaction is in
general one or two orders of magnitude smaller when compared to the isotropic
one. For the magnetic properties of Ni clusters the magnetic properties can
show quite a different behaviour and we find in this case a strong tendency
towards noncollinear magnetism.
\end{abstract}

\maketitle

\section{Introduction}

The magnetism of surface supported clusters has been the subject of intense
research activities over the last few years as such systems often show peculiar
and unexpected magnetic behaviour. These exceptional magnetic properties arise
from the reduced dimensionality in combination with spin-orbit coupling (SOC)
which can cause complex interactions among the atomic magnetic moments. In this
context clusters of magnetic 3{\it d} transition metal elements deposited on
5{\it d} noble metal substrates are very interesting as for these systems
spin-orbit driven effects mediated by substrate atoms with large SOC are most
prominent. With technical or chemical applications in focus, there is a growing
need to understand the trends and principles behind the manifold of magnetic
properities for different cluster and substrate materials as only  this will
make it possible to anticipate which magnetic properties may result from a
particular cluster/substrate combination.

In previous experimental and theoretical investigations on the magnetism of
atomic clusters on surfaces it was already demonstrated that their magnetic
properties differ strongly from the magnetic properties of the corresponding
bulk materials and that this has its main origin in the reduced atomic
coordination of cluster sites which in fact has a strong impact on the local
spin and orbital magnetic moments~\cite{GRV+03,SBM+07,MLB10}.  More recently it
was also shown that for 3{\it d} clusters or monolayers on 5{\it d} metal
surfaces SOC induced effects on the spin configurations also play an immanent
role causing various noncolliniear magnetic
structures~\cite{ALU+08,RGDP11,MLB+10}. This SOC induced noncolliniear
magnetism is, however, intrinsically different from the spin frustrations that
may arise e.g. by a competition between ferro- and anti-ferromagnetism or that
may be present in systems where the magnetic and geometric symmetries are
incompatible~\cite{LMZ+07,RRDP10}.

Unfortunately, each of the theoretical studies published so far were aimed at
only one or two combinations of the cluster and substrate materials and often
only very few cluster sizes and shapes were investigated. In addition to that
comes the fact that many theoretical investigations have focused only on some
selected magnetic properties as for instance the magnetic moments and exchange
interaction but leaving out important information concerning the magnetic
anisotropy energy (MAE).  Moreover, due to limitations which are present in all
theoretical schemes it is often also problematic to compare results obtained
for different systems by different groups which use different methods.  Thus,
in order to obtain a more complete picture about the trends in the magnetic
properties of deposited clusters one needs a sufficiently large self-contained
set of results for interrelated systems which are obtained by the same method.
This motivated us to calculate a large spectrum of the magnetic properties for
sets of Fe, Co, and Ni clusters of 1-7 atoms on Ir(111), Pt(111), and Au(111)
surfaces, within a unified fully-relativistic Green's function formalism.
Moreover, we studied also complete monolayers as reference systems for the
sequences with increasing cluster size.  This enables us to analyse a large
pool of data which are directly comparable because they were obtained by the
same procedure.  We found that the magnetism of Fe and Co  clusters on all
investigated surfaces follows common patterns that can be understood by
considering the coordination numbers of atoms in the clusters and the
polarisability of the substrate.  For Ni clusters the situation is more
complicated and some of the systematic trends observed for Fe and Co clusters
are absent.

\section{Computational framework}

The calculations for the investigated cluster and monolayer systems were done
within the framework of spin density functional theory (SDFT) using the local
spin density approximation (LSDA) with the parametrisation given by Vosko, Wilk
and Nusair for the exchange and correlation potential~\cite{VWN80}. The
electronic structure has been determined in a fully relativistic way on the
basis of the Dirac equation for spin-polarised potentials which was solved
using the Korringa-Kohn-Rostoker (KKR) multiple scattering
formalism~\cite{EKM11}.  The calculations for surface deposited clusters
consist of two steps. First the host surface is calculated self-consistently
with the tight-binding or screened version of the KKR method~\cite{ZDU+95}
using layers of empty sites to represent the vacuum region. This step is then
followed by treating the deposited clusters as a perturbation to the clean
surface with the Green's function for the new system being obtained by solving
the corresponding Dyson equation~\cite{BMP+05}.  This technique avoids the
spurious interactions between clusters which may occur if a supercell approach
is used instead~\cite{SBME10}.

For all systems discussed below the cluster atoms were assumed to occupy ideal
lattice sites in the first vacuum layer and no effects of structure relaxation
were included. The substrates were simulated by finite slabs which contained 37
atomic layers  and we used lattice parameters of 3.84~\AA, 3.92~\AA\ and
4.08~\AA\ for Ir(111), Pt(111) and Au(111), respectively.  The surface
calculations were converged with respect to $\vec k$-point integration. For the
surface Brillouin zones a regular $\vec k$-mesh of $100\times100$ points was
used which corresponds to 1717 $\vec k$-points in the irreducible part of the
Brillouin zone.  The effective potentials were treated within the atomic sphere
approximation (ASA).  The occuring energy integrals were evaluated by contour
integration on a semicircular path within the complex energy plane using a
logarithmic mesh of 32 points. The multipole expansion of the Green's function
was truncated at an angular momentum cutoff of $\ell_{\rm max} = 2$. For
selected surface and cluster systems calculations with $\ell_{\rm max} = 3$
were also performed which showed that this causes a more-or-less uniform
increase of the local spin moments by 3-5\% and of the local orbital moments by
3-10\%. This indicates that the systematic trends in the spin and orbital
magnetic moments are well described by $\ell_{\rm max} = 2$.
 
For the representation of the interatomic exchange interactions we made use of
the rigid spin approximation~\cite{AKH+96} and mapped the magnetic energy
landscape $E(\{\hat{e}_k\})$ onto an extended classical Heisenberg model for
all atomic magnetic moment directions $\{\hat{e}_k\}$. The corresponding
extended Heisenberg Hamiltonian has the form~\cite{AKL97,USPW03}: 
%%%%%%%%%%%%%%%%%%%%%%%%%%%%%%%%%%%%%%%%%%%%%%%%%%%%%%%%%%%%%%%%%%%%%%%%%
\begin{eqnarray} 
H &=& -\frac{1}{2}\sum_{i,j (i \neq j)} J_{ij} \hat{e}_i \cdot \hat{e}_j 
      -\frac{1}{2}\sum_{i,j (i \neq j)} \hat{e}_i {\mathcal J}^S_{ij}
       \hat{e}_j \nonumber \\ 
  & & -\frac{1}{2}\sum_{i,j (i \neq j)} 
       \vec{D}_{ij}\cdot [\hat{e}_i\times \hat{e}_j] 
      +\sum_{i}  K_i (\hat{e}_i)\;,
\label{Hspin_2} 
\end{eqnarray}
%%%%%%%%%%%%%%%%%%%%%%%%%%%%%%%%%%%%%%%%%%%%%%%%%%%%%%%%%%%%%%%%%%%%%%%%%
where, the exchange interaction tensor has been decomposed into its
conventional isotropic part $J_{ij}$, its traceless symmetric part ${\mathcal
J}^S_{ij}$ and its anti-symmetric part ${\mathcal J}^A_{ij}$ which is given in
terms of the Dzyaloshinski-Moriya (DM) vector $\vec{D}_{ij}$.  We calculated
the $J_{ij}$ coupling parameters and DM vectors $\vec{D}_{ij}$ following the
scheme by Udvardi et al.~\cite{USPW03}.

The anisotropy constants $K_i(\hat{e}_i)$ account for the on-site magnetic
anisotropy energy associated with each individual magnetic moment oriented
along $\hat{e}_i$.   The magnetic anisotropy energy $\Delta E$ is usually split
into two parts, the SOC induced magnetocrystalline anisotropy $\Delta
E_{\text{soc}}$ and the so-called shape anisotropy $\Delta E_{\text{dd}}$
caused by magnetic dipole-dipole interactions, i.e.
%%%%%%%%%%%%%%%%%%%%%%%%%%%%%%%%%%%%%%%%%%%%%%%%%%%%%%%%%%%%%%%%%%%%%%%%%
\begin{eqnarray} 
\Delta E &=& \Delta E_{\text{soc}} + \Delta E_{\text{dd}} \;.
\label{mae} 
\end{eqnarray}
%%%%%%%%%%%%%%%%%%%%%%%%%%%%%%%%%%%%%%%%%%%%%%%%%%%%%%%%%%%%%%%%%%%%%%%%%
$\Delta E_{\text{dd}}$ can be determined classically by a lattice summation
over the magnetostatic energy contributions of the individual magnetic moments
or in an ab-initio way by using a Breit Hamiltonian~\cite{BMB+12}.  Here, we
used the classical approach to calculate $\Delta E_{\text{dd}}$ for the full
monolayers while we found that for clusters containing just a few magnetic
atoms $\Delta E_{\text{dd}}$ is negligible. The magnetocrystalline anisotropy
energy $\Delta E_{\text{soc}}$ was extracted from magnetic torque calculations
which are described in more detail in Refs.~\cite{MBM+09,SBME10,SSB+06}.

As discussed recently by \v{S}ipr et al.~\cite{SBME10} the approximations and
truncations mentioned in this Section result in a limited accuracy  concerning
in particular the values of $\Delta E_{\text{soc}}$.  However, this does not
hinder our analysis of the general trends of $\Delta E_{\text{soc}}$ with
respect to cluster geometries as well as different cluster/substrate
combinations.

\section{Results and Discussion}

\subsection{Magnetic moments}
%%%%%%%%%%%%%%%%%%%%%%%%%%%%%%%%%%%%%%%%%%%%%%%%%%%%%%%%%%%%%%%%%%%%%%%%%%%%%%%%
\begin{figure*}
\includegraphics[width=2.00\columnwidth,clip]{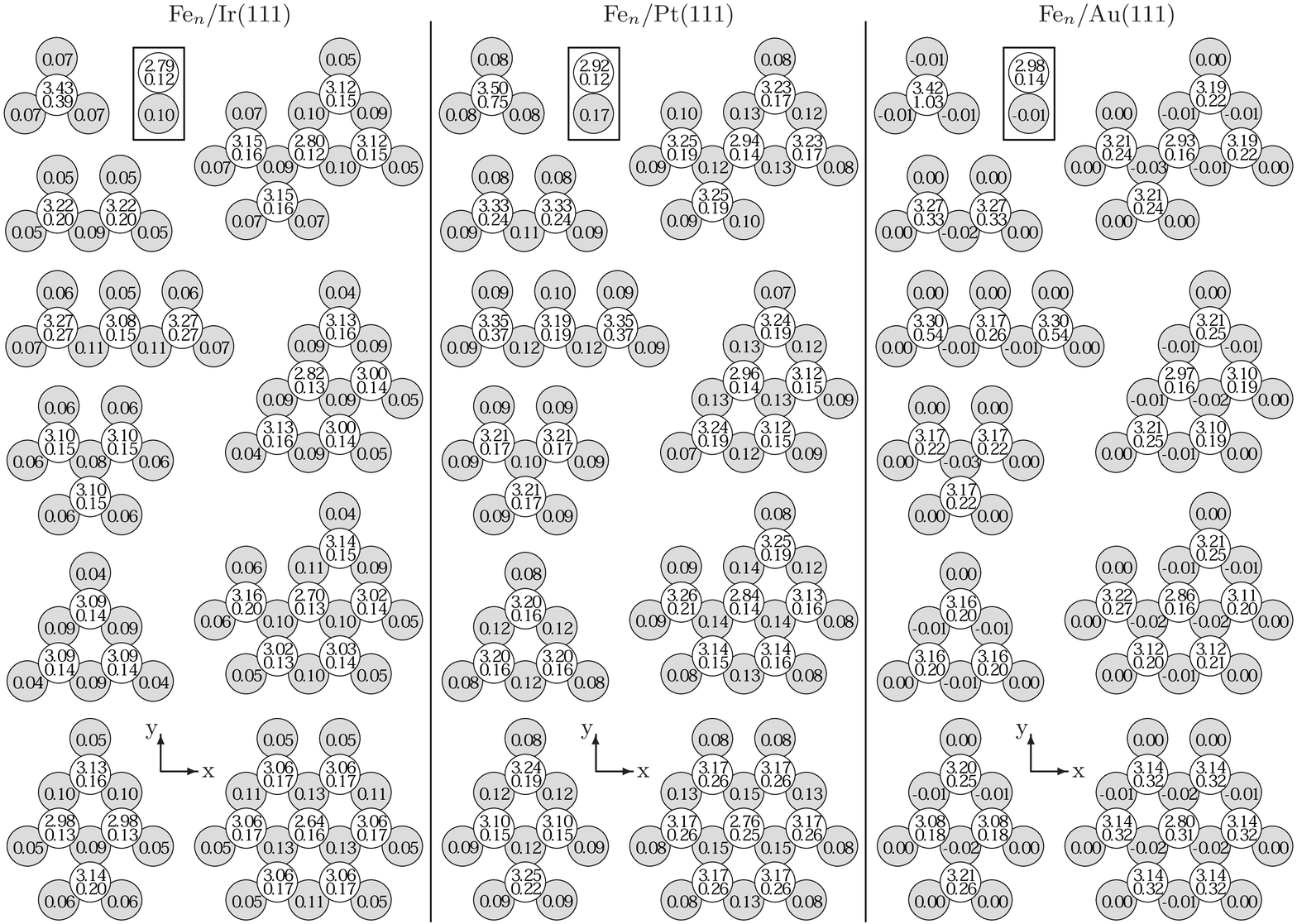}
\caption{\label{spin_cluFe} Cluster geometries for Fe clusters of 1-7 atoms
supported by Ir(111), Pt(111) and Au(111).  The local spin and orbital magnetic
moments at Fe sites are given by the upper and lower numbers, respectively. The
spin magnetic moments for nearest neighbour substrate sites are also shown.
The data presented within frames give the corresponding monolayer values.}
\end{figure*}
%%%%%%%%%%%%%%%%%%%%%%%%%%%%%%%%%%%%%%%%%%%%%%%%%%%%%%%%%%%%%%%%%%%%%%%%%%%%%%%%
%%%%%%%%%%%%%%%%%%%%%%%%%%%%%%%%%%%%%%%%%%%%%%%%%%%%%%%%%%%%%%%%%%%%%%%%%%%%%%%%
\begin{figure*}
\includegraphics[width=2.00\columnwidth,clip]{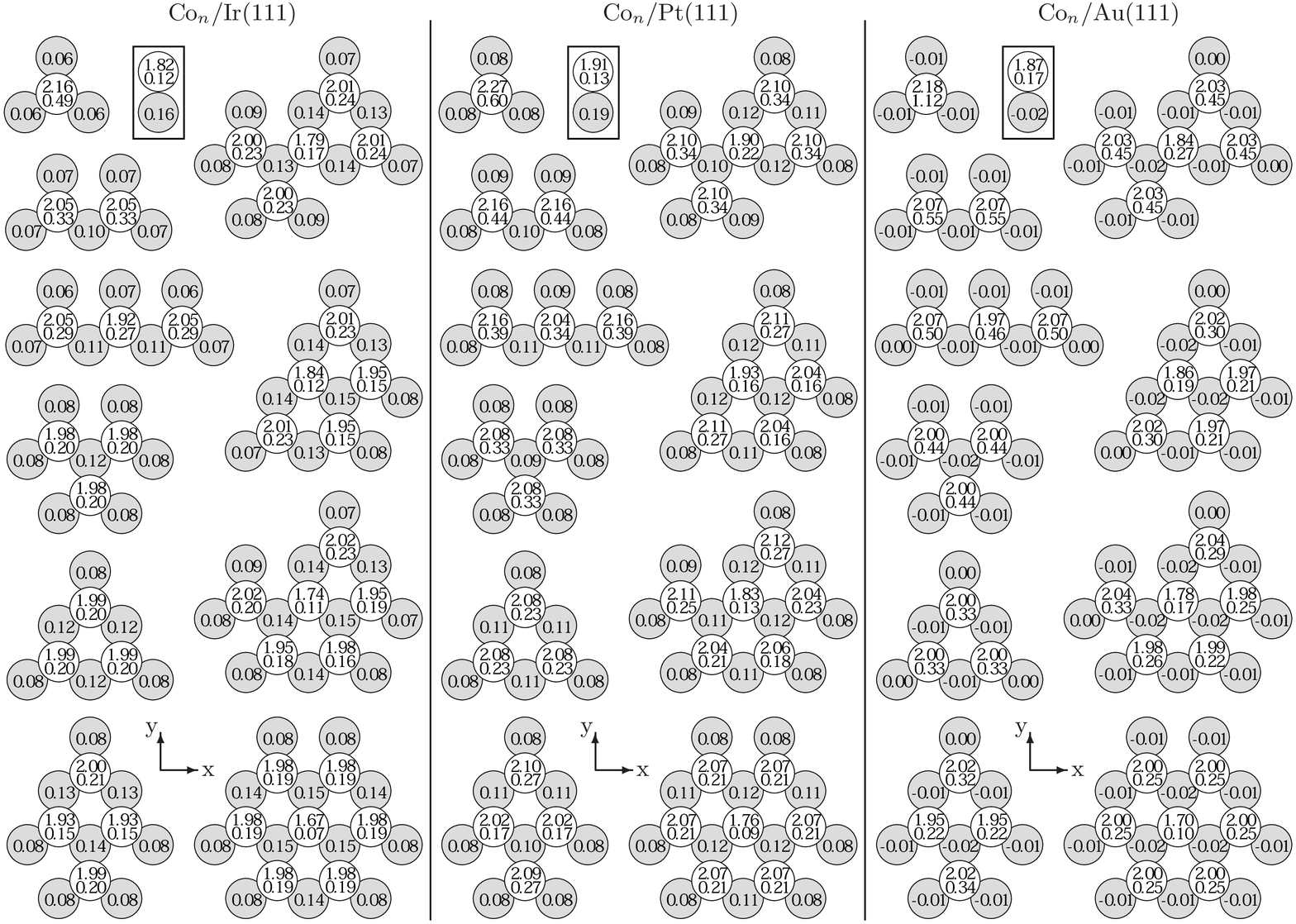}
\caption{\label{spin_cluCo} Cluster geometries for Co clusters of 1-7 atoms
supported by Ir(111), Pt(111) and Au(111).  The local spin and orbital magnetic
moments at Co sites are given by the upper and lower numbers, respectively. The
spin magnetic moments for nearest neighbour substrate sites are also shown.
The data presented within frames give the corresponding monolayer values.}
\end{figure*}
%%%%%%%%%%%%%%%%%%%%%%%%%%%%%%%%%%%%%%%%%%%%%%%%%%%%%%%%%%%%%%%%%%%%%%%%%%%%%%%%
%%%%%%%%%%%%%%%%%%%%%%%%%%%%%%%%%%%%%%%%%%%%%%%%%%%%%%%%%%%%%%%%%%%%%%%%%%%%%%%%
\begin{figure*}
\includegraphics[width=2.00\columnwidth,clip]{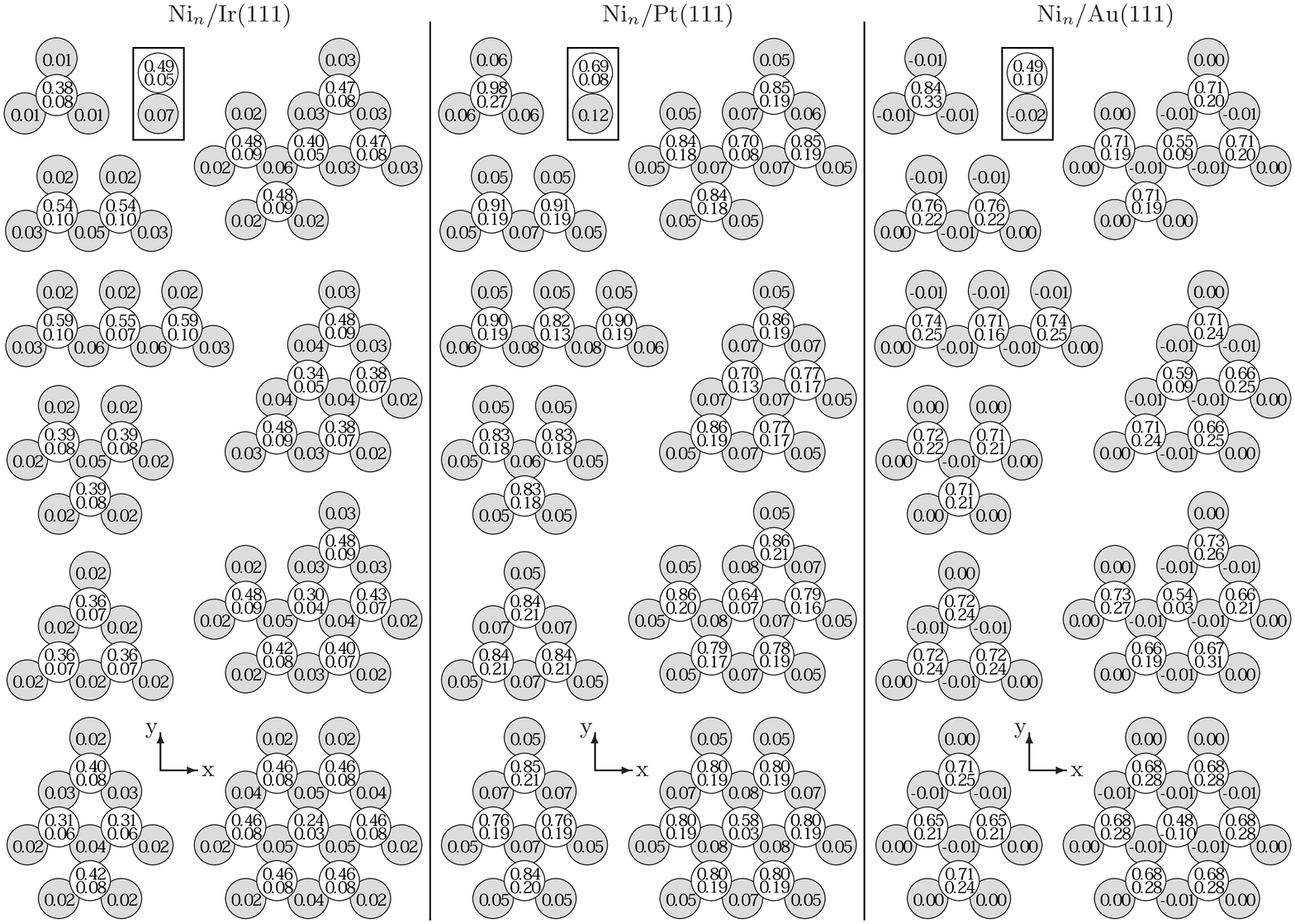}
\caption{\label{spin_cluNi} Cluster geometries for Ni clusters of 1-7 atoms
supported by Ir(111), Pt(111) and Au(111).  The local spin and orbital magnetic
moments at Ni sites are given by the upper and lower numbers, respectively. The
spin magnetic moments for nearest neighbour substrate sites are also shown.
The data presented within frames give the corresponding monolayer values.}
\end{figure*}
%%%%%%%%%%%%%%%%%%%%%%%%%%%%%%%%%%%%%%%%%%%%%%%%%%%%%%%%%%%%%%%%%%%%%%%%%%%%%%%%
Fig.~\ref{spin_cluFe} shows the considered cluster geometries together with the
calculated values of local spin ($\mu_{\mathrm{spin}}$) and orbital
($\mu_{\mathrm{orb}}$) magnetic moments for Fe clusters of 1-7 atoms as well as
full Fe monolayers deposited on Ir(111), Pt(111) and Au(111). For identical Co
and Ni clusters the corresponding data are presented in Figs.~\ref{spin_cluCo}
and \ref{spin_cluNi}.  In addition, these figures also show the induced spin
magnetic moments of the respective substrate atoms that are adjacent to cluster
atoms.  One can see that in some cases there are considerable variations of
$\mu_{\mathrm{spin}}$ and $\mu_{\mathrm{orb}}$ between the different sites of
the deposited clusters.  The magnetic moments depend not only on the position
of the site with respect to other Fe, Co or Ni atoms but also on their position
with respect to the underlying substrate atoms.  This can be observed for
example when inspecting the two differently located compact trimers or the
cross-shaped five atom clusters in Figs.~\ref{spin_cluFe},~\ref{spin_cluCo}
and~\ref{spin_cluNi}.  Clusters supported by Pt(111) have largest
$\mu_{\mathrm{spin}}$ when compared with Ir(111) and Au(111) while the
$\mu_{\mathrm{orb}}$ values are increasing from Ir to Pt to Au.

There is a big difference between the induced magnetic moments in the Ir(111)
and Pt(111) substrates on the one hand and the Au(111) substrate on the other
hand.  Ir and Pt atoms which are nearest neighbours of any Fe or Co atom have a
relatively large $\mu_{\mathrm{spin}}$ of up to 0.15 $\mu_{\mathrm{B}}$, while
corresponding Au atoms  have always small negative $\mu_{\mathrm{spin}}$, not
larger than 0.03 $\mu_{\mathrm{B}}$ in the absolute value. Substrate atoms with
a larger number of Fe, Co or Ni neighbours usually have a larger
$\mu_{\mathrm{spin}}$ than substrate atoms with a smaller number of
neighbouring cluster atoms.  However, this is not a general rule as seen for
the substrate atoms adjacent to the central atom of the cross-shaped Fe$_5$ and
Co$_5$ or to the differently located compact Fe$_3$ and Co$_3$ clusters on
Ir(111) and Pt(111).

The orbital magnetic moments induced in the substrate atoms are always small:
they can reach up to 0.03~$\mu_{\mathrm{B}}$ for Fe and Co on Pt(111) while
being smaller than 0.007~$\mu_{\mathrm{B}}$ for Ir(111) and smaller than
0.004~$\mu_{\mathrm{B}}$ for Au(111). Except for the Au(111) substrate atoms
$\mu_{\mathrm{orb}}$ is found to be always parallel with $\mu_{\mathrm{spin}}$.
The finding that Pt is the most polarisable of the three elements and that Au
is less polarisable than Ir is consistent with earlier
theoretical~\cite{GE95,TvdLT+03} and experimental
works~\cite{WPC+00,WPW+01,WAJ+04,KSK+06} for multilayer systems.  This high
spin polarisability of Pt can be ascribed to its high spin susceptibility that
in turn is caused by its relatively large density of states at the Fermi level
leading to a large Stoner product (see below).

\begin{figure*}
\includegraphics[width=2.00\columnwidth,clip]{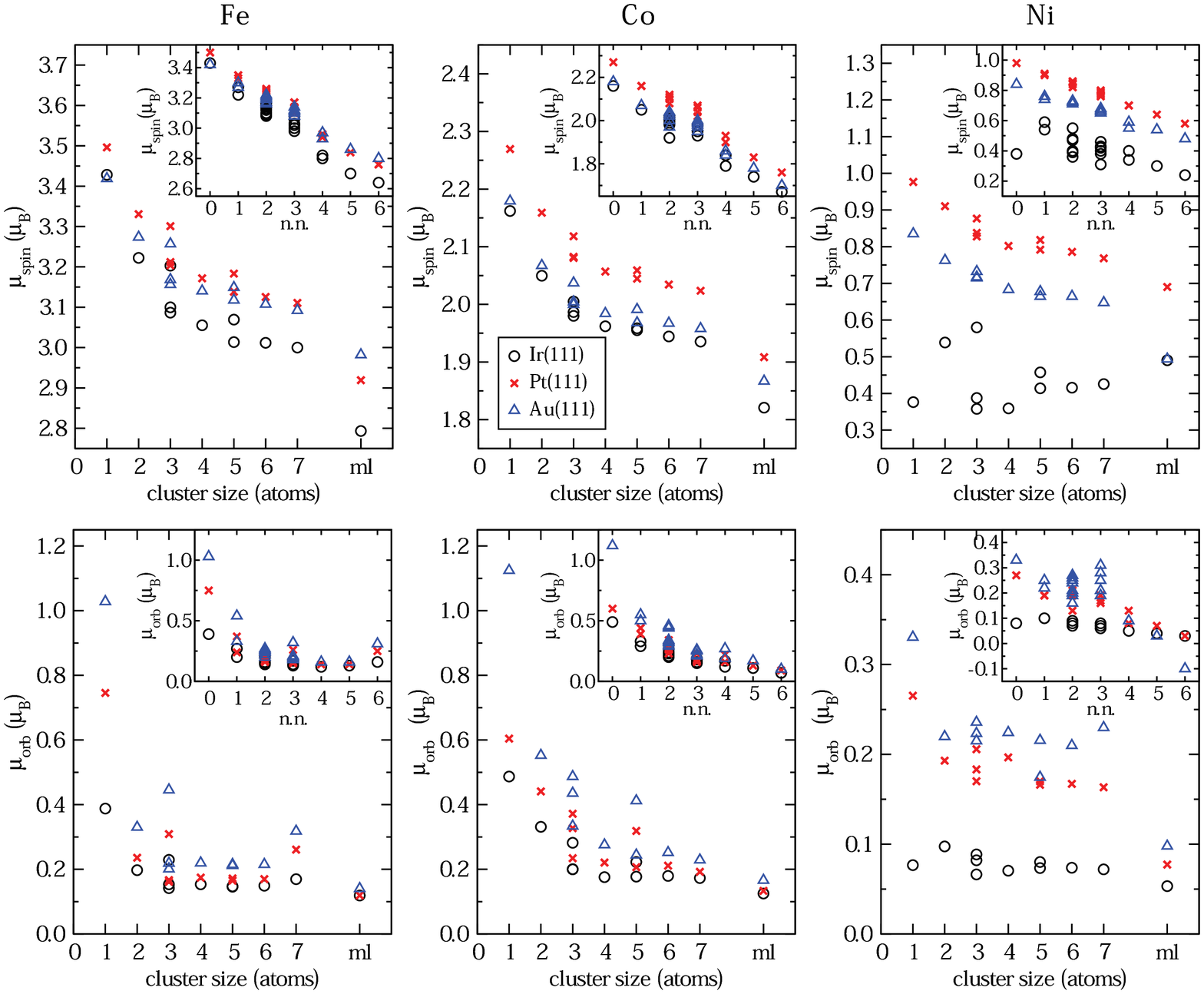}
\caption{\label{average_spin}
Average spin (top row) and orbital (bottom row) magnetic moments per atom for Fe,
Co and Ni clusters and monolayers (ml) on Ir(111), Pt(111) and Au(111),
respectively. The insets present $\mu_{\mathrm{spin}}$ and $\mu_{\mathrm{orb}}$
vs. the number of nearest neighbouring cluster atoms (n.n.).}
\end{figure*}

By plotting the local magnetic moments as a function of the coordination number
one can visualise the site-dependence of $\mu_{\mathrm{spin}}$ and
$\mu_{\mathrm{orb}}$.  Such plots are shown in the insets of
Fig.~\ref{average_spin}, where only neighbouring cluster atoms are considered
in defining the coordination number. Sites with a lower coordination number
generally have larger $\mu_{\mathrm{spin}}$ and $\mu_{\mathrm{orb}}$ than sites
with a higher coordination number, with Ni on Ir(111) being the only exception.
For Ni on Ir(111) the magnetic moments quasi-oscillate strongly with changing
cluster size or shape and we find that $\mu_{\mathrm{spin}}$ for the adatom
(0.38~$\mu_{\mathrm{B}}$) is smaller than for a Ni atom in the full monolayer
(0.49~$\mu_{\mathrm{B}}$).  For all other cluster/substrate systems considered
in our study a quasi-linear relationship between $\mu_{\mathrm{spin}}$ and
coordination number is found.  Interestingly, increasing the coordination
number for the atoms of such small clusters leads to a stronger reduction of
$\mu_{\mathrm{spin}}$ when compared to equally coordinated atoms in larger
clusters or full monolayers.  For example the central atom of a compact 7-atom
cluster has always lower $\mu_{\mathrm{spin}}$ than a monolayer atom.  One can
also see that the corresponding orbital magnetic moments are much more
sensitive with respect to coordination than the spin magnetic moments.  While
the insets in Fig.~\ref{average_spin} show a strong decay of
$\mu_{\mathrm{orb}}$ with increasing coordination for Fe and Co clusters on all
three substrates the orbital magnetism in Ni clusters behaves non-monotonously.

%%%%%%%%%%%%%%%%%%%%%%%%%%%%%%%%%%%%%%%%%%%%%%%%%%%%%%%%%%%%%%%%%%%%%%%%%%%%%
%%%%%%%%%%%%%%%%%%%%%%%%%%%%%%%%%%%%%%%%%%%%%%%%%%%%%%%%%%%%%%%%%%%%%%%%%%%%%
An analysis of the average spin and orbital magnetic moments as function of
cluster size is shown in Fig.~\ref{average_spin}. For the three and five atom
clusters the lower $\mu_{\mathrm{spin}}$ and $\mu_{\mathrm{orb}}$ values
correspond to the compact clusters. All clusters have largest
$\mu_{\mathrm{spin}}$ when deposited on Pt(111) followed by the Au(111)
substrate. The lowest $\mu_{\mathrm{spin}}$ values are obtained for deposition
on Ir(111). The highest values of $\mu_{\mathrm{orb}}$, however, are found for
clusters deposited on Au(111) where the interaction between cluster and
substrate atoms is weak and the lattice constant largest.  

Concerning the trend of $\mu_{\mathrm{spin}}$ for the three different
substrates there are two competing effects that must be considered. At first
there is an increase in the lattice constant when going from Ir (a=3.84~\AA) to
Pt (a=3.92~\AA) to Au (a=4.08~\AA), i.e.\ as the atoms of the clusters occupy
ideal lattice sites, their distance from the substrate is largest in the case
of Au.  This means also that among the discussed substrates, the interaction
between adatoms and substrate is smallest for Au and one would therefore expect
that clusters deposited on Au(111) would have the largest $\mu_{\mathrm{spin}}$
values. On the other hand hybridisation of the electronic states between
adatoms and substrate leads to a small charge transfer of minority
3$d$-electrons from the cluster atoms into empty 5$d$-states of adjacent
substrate atoms, thereby increasing $\mu_{\mathrm{spin}}$ for the clusters.
This however, happens only for the spatially extended 5$d$-states of Ir and Pt
with their 5$d$-states having an appreciable energetic overlap with the
minority 3$d$-states of cluster atoms.  This can be clearly seen from the
density of states curves which are presented in Fig.~\ref{dosFeCoNi}. In
contrast to this there is almost no or only little interaction between the
minority 3$d$-states of cluster atoms with the energetically low-lying
5$d$-states of Au. Besides, the hybridization between the cluster-derived and
substrate-derived states leads to energy lowering of the Fe, Co and Ni
4$p$-states and this lowering is again more pronounced for the Ir and Pt
substrates than for the Au substrate.

This causes an additional charge redistribution within the cluster atoms, i.e.\
4$p$-states become occupied, again at the cost of minority 3$d$-states. In this
way Fe$_1$ deposited on Pt(111) ends up with about 0.1 electrons less in the
minority 3$d$-orbitals and thus a slightly larger spin magnetic moment when
compared to the deposition on Au(111), with this effect becoming stronger for
Co and Ni.  \v{S}ipr et al.~\cite{SBM+07} have demonstrated this behaviour for
Co clusters on Pt(111) and Au(111)  by performing calculations for Co clusters
on an Au(111) substrate using the lattice constant of Pt. This also showed that
the observed increase in $\mu_{\mathrm{orb}}$ can be solely attributed to the
larger lattice constant of Au.

\subsection{Density of states}

%%%%%%%%%%%%%%%%%%%%%%%%%%%%%%%%%%%%%%%%%%%%%%%%%%%%%%%%%%%%%%%%%%%%%%%%%%%%%%%%
\begin{figure*}
\includegraphics[width=2.00\columnwidth,clip]{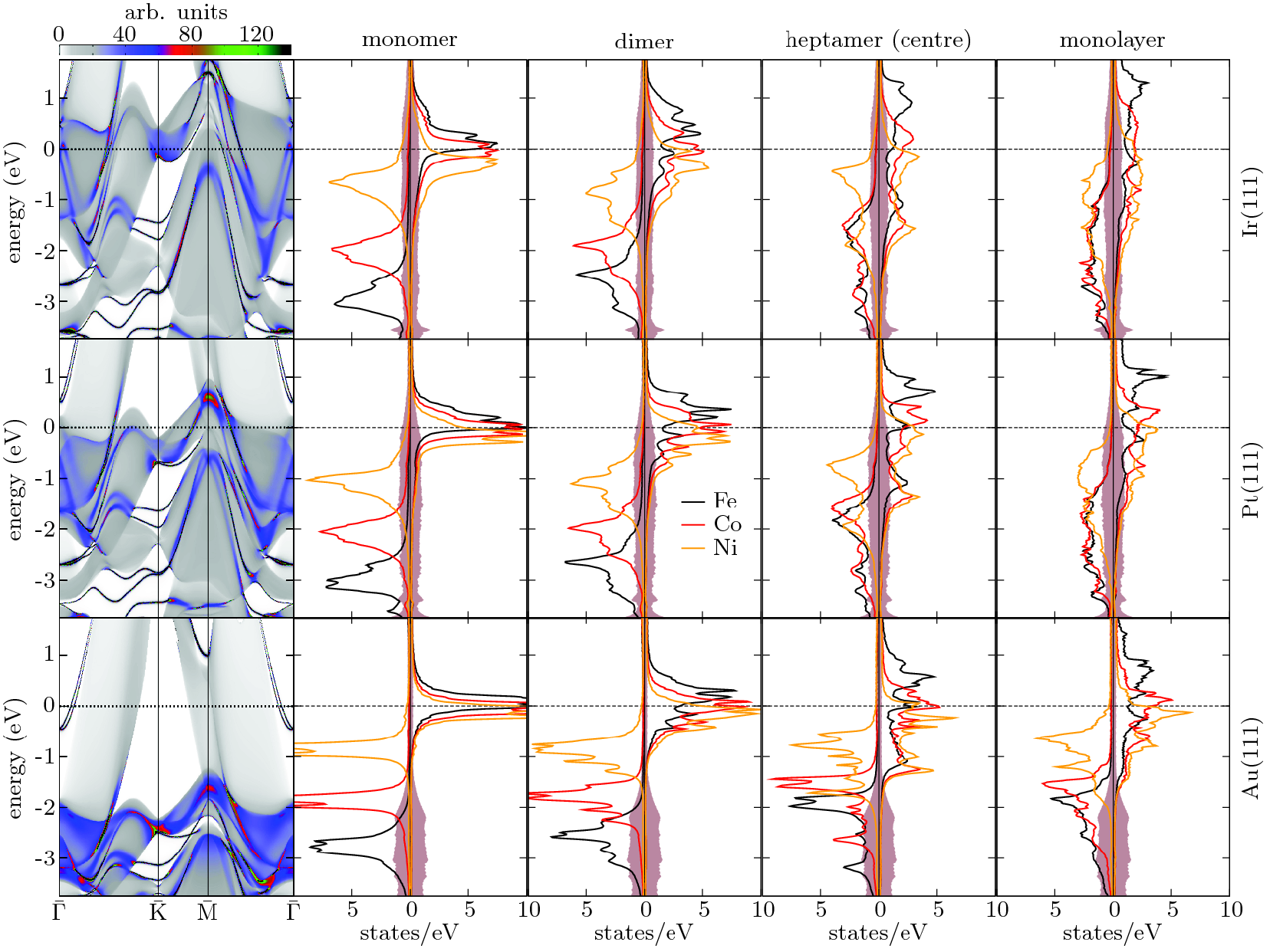}
\caption{\label{dosFeCoNi} Spin projected density of states (DOS) for Fe, Co
and Ni monomers (second column), dimers (third column) as well as the central
atom of a 7 atom cluster (fourth column) and the corresponding full monolayers
(rightmost column) deposited on the (111) surfaces of Ir (top row), Pt (middle
row) and Au (bottom row).  The brown areas represent the DOS for unperturbed
surface atoms of clean substrates.  Corresponding  Bloch spectral functions
$A^B(\vec{k},E)$ for surface atoms of clean substrates are presented in the
leftmost column along the $\bar{\Gamma}$-$\bar{\rm K}$-$\bar{\rm M}$-$\bar{\Gamma}$ 
line of the two-dimensional Brillouin zone.}
\end{figure*}
%%%%%%%%%%%%%%%%%%%%%%%%%%%%%%%%%%%%%%%%%%%%%%%%%%%%%%%%%%%%%%%%%%%%%%%%%%%%%%%%
In Fig.~\ref{dosFeCoNi} we show the spin-resolved density of states (DOS) for
the adatoms, dimers and the central atoms of the 7-atom clusters as well as the
corresponding full  monolayers. The DOS of the respective topmost atomic layer
of the clean substrates is also shown by the brown areas together with the
respective Bloch spectral function $A^B(\vec{k},E)$ along the high symmetry
lines $\bar{\Gamma}$-$\bar{\rm K}$-$\bar{\rm M}$-$\bar{\Gamma}$ in the
two-dimensional Brillouin zone presented in the first column. $A^B(\vec{k},E)$
can be interpreted as a $\vec{k}$-resolved DOS revealing the detailed features
in the electronic structure of the three different substrates.  The large
grey-shaded regions in the $A^B(\vec{k},E)$ diagrams in the first column of
Fig.~\ref{dosFeCoNi} correspond to electronic bulk states of the underlying
substrates, while the sharp black lines represent surface states localised
within the topmost atomic layer of the clean surfaces. The blue and red regions
arise from hybridisations between surface and bulk states.  Clearly visible are
for instance the Rashba split surface states around $\bar{\Gamma}$ for Pt(111)
and Au(111).

For Ir and Pt there is an appreciable energetic overlap between electronic
states located at the substrate and states located at cluster sites resulting
in hybridisation with a prominent broadening in the cluster DOS.  The situation
is different for the Au substrate,  where the energetically low-lying states of
Au can only hybridise with the majority states of Fe while for minority states
of Fe as well as for all states of Co and Ni, there are no energetically close
Au states to hybridize with and hence very distinct atomic-like features
prevail in the DOS of cluster atoms in that case.  With increasing number of
cluster atoms a complex fine structure appears in the DOS which also broadens
appreciably with increasing coordination numbers of the cluster atoms.
Moreover, the presented DOS curves for the central atom of the 7-atom cluster
demonstrate that the DOS of deposited clusters acquires very quickly the main
features which are present in the DOS of a complete monolayer.  However, as for
such small clusters the DOS at the Fermi level varies strongly with changing
the number of atoms so do the corresponding chemical and magnetic properties in
this finite size regime.

The decreasing overlap between states located in the substrate and located in
the clusters when going from Ir to Au explains the finding that $\mu_{\rm orb}$
is largest for clusters on Au where the atomic-like character of the DOS
prevails.  In this context, however, the size of $\mu_{\rm spin}$ cannot always
be directly related to the overlap between cluster and substrate DOS as this
overlap is smaller for  Au(111) than for Pt(111) and yet $\mu_{\rm spin}$ is
largest for clusters on the Pt(111) substrate.

In the same way also the induced magnetisation within the substrate depends on
this mutual energetic overlap between 3$d$ and 5$d$-states, which explains the
very small induced magnetic moments in the case of Au(111).  But also here, one
should keep in mind that the polarizability of the substrate atoms is
determined by the Stoner product $I \cdot N_{F}$ with the exchange integral $I$
and the number of states at the Fermi level $N_{F}$ ($I \cdot N_{F}$ = 0.29 for
Ir, 0.59 for Pt and 0.05 for Au)~\cite{SP94}.

\subsection{Exchange Coupling}

%%%%%%%%%%%%%%%%%%%%%%%%%%%%%%%%%%%%%%%%%%%%%%%%%%%%%%%%%%%%%%%%%%%%%%%%%%%%%
\begin{table}%[b]
\caption{Isotropic exchange coupling constants $J_{ij}$ in meV for Fe, Co and Ni clusters
         deposited on Ir(111), Pt(111) and Au(111). The icons in the left column
         indicate the corresponding cluster geometry as well as the cluster sites
         $i$ and $j$, respectively. The last line gives $J_{ij}$ for nearest neighbouring
         sites within the full monolayer (ml).}
\label{exchange_coupling}
\begin{center}
\begin{ruledtabular}
\begin{tabular}{crrrrrrrrrr}

&\multicolumn{3}{c}{Fe}&\multicolumn{3}{c}{Co}&\multicolumn{3}{c}{Ni} \\
%\cmidrule(r){2-4} \cmidrule(r){5-7} \cmidrule(r){8-10}
$ij$    &\multicolumn{1}{c}{Ir}&\multicolumn{1}{c}{Pt}&\multicolumn{1}{c}{Au}
        &\multicolumn{1}{c}{Ir}&\multicolumn{1}{c}{Pt}&\multicolumn{1}{c}{Au}
        &\multicolumn{1}{c}{Ir}&\multicolumn{1}{c}{Pt}&\multicolumn{1}{c}{Au} \\
%\midrule
\includegraphics[height=10pt]{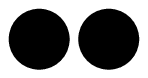}    &128.8&137.8&143.8& 97.5&107.7&112.8&  7.9& 30.4& 26.7  \\
\includegraphics[height=10pt]{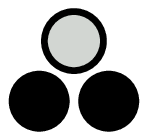}  &110.5&111.8&114.6& 69.8& 77.6& 72.5&  0.8& 14.6& 16.1  \\
\includegraphics[height=10pt]{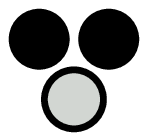}  &100.3&107.9&114.2& 64.5& 71.0& 67.4&  1.8& 16.4& 15.7  \\
\includegraphics[height=10pt]{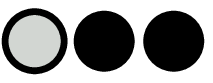}   & 76.8& 90.9& 90.6& 66.9& 76.9& 74.8& 11.2& 24.9& 20.9  \\
\includegraphics[height=10pt]{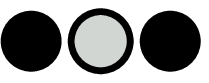}   &-15.9& -8.4&-12.2&  5.2&  3.9& -0.4&  4.3&  6.1&  3.4  \\
\includegraphics[height=10pt]{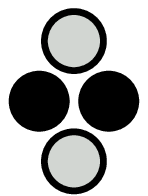}    & 79.5& 79.2& 82.3& 49.5& 59.8& 47.5&  0.4&  7.6&  8.0  \\
\includegraphics[height=10pt]{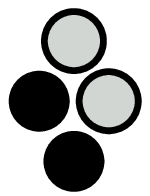}    & 83.0& 92.1& 99.4& 59.3& 66.1& 60.5&  2.0& 15.9& 13.5  \\
\includegraphics[height=10pt]{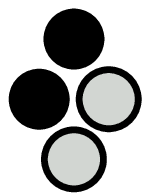}    & 97.2& 99.4&100.7& 61.4& 69.7& 60.2&  1.4& 14.6& 12.7  \\
\includegraphics[height=10pt]{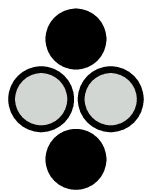}    & -1.2& -0.8& -3.8&  6.7&  8.4& 10.1&  0.4&  3.3&  1.6  \\
\includegraphics[height=10pt]{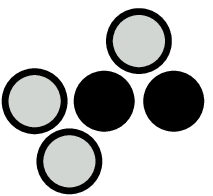} & 79.6& 79.2& 77.0& 53.4& 61.5& 55.4&  3.8& 15.1& 10.8  \\
\includegraphics[height=10pt]{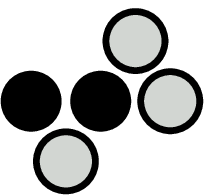} & 51.2& 64.9& 67.0& 48.4& 54.0& 46.3&  4.2& 15.1& 10.2  \\
\includegraphics[height=10pt]{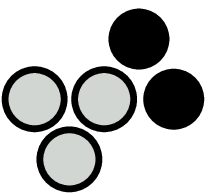} &118.3&117.4&122.2& 65.7& 69.3& 65.1&  1.0& 16.1& 14.0  \\
\includegraphics[height=10pt]{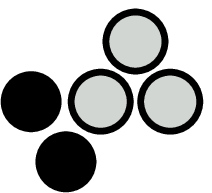} &109.6&110.2&116.3& 66.3& 72.2& 70.0&  2.8& 16.7& 14.2  \\
\includegraphics[height=10pt]{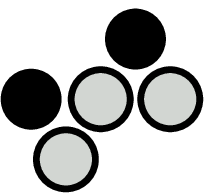} &  0.7&  4.2&  5.9&  1.5&  4.2&  6.9& -0.6&  1.5&  2.2  \\
\includegraphics[height=10pt]{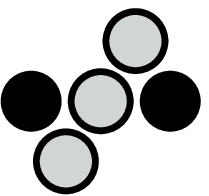} &-12.4& -7.6&-10.4&  4.7&  6.2&  4.0&  2.2&  4.9&  3.5  \\
\includegraphics[height=10pt]{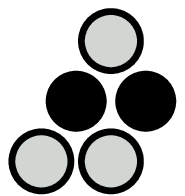}    & 74.2& 75.6& 75.9& 50.3& 57.9& 45.1&  0.8&  7.9&  7.1  \\
\includegraphics[height=10pt]{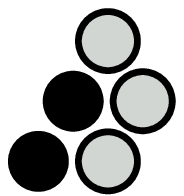}    & 71.8& 79.3& 68.8& 48.6& 53.8& 46.2&  3.8& 16.9& 11.7  \\
\includegraphics[height=10pt]{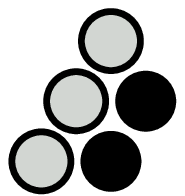}    & 67.8& 77.1& 89.8& 45.1& 56.0& 44.8&  3.3& 15.1& 10.0  \\
\includegraphics[height=10pt]{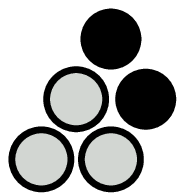}    &104.0&105.5&105.4& 64.0& 73.0& 65.9&  2.1& 14.4&  9.6  \\
\includegraphics[height=10pt]{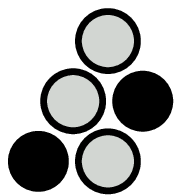}    &  1.2&  0.3& -1.9&  3.1&  7.9&  9.4&  0.1&  2.4&  0.9  \\
\includegraphics[height=10pt]{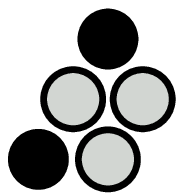}    & -6.3& -4.4& -4.3&  4.0&  4.6&  5.4&  2.1&  5.4&  3.5  \\
\includegraphics[height=10pt]{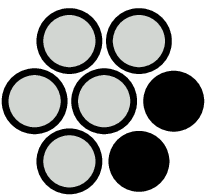}    & 61.0& 72.9& 87.9& 48.2& 59.1& 50.8&  3.9& 13.1&  8.8  \\
\includegraphics[height=10pt]{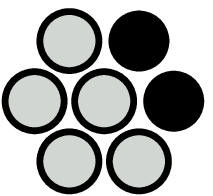}    & 88.8& 88.1& 89.9& 50.6& 59.4& 45.0&  3.4& 13.1&  8.6  \\
\includegraphics[height=10pt]{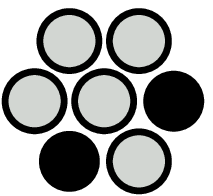}    &  2.4&  1.6&  2.2&  1.0&  5.3&  7.8&  0.4&  1.8&  0.1  \\
\includegraphics[height=10pt]{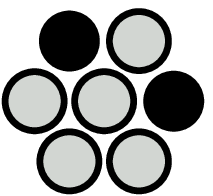}    &  2.4&  1.6&  2.2&  1.0&  5.3&  7.8&  0.4&  1.8&  0.1  \\
\includegraphics[height=10pt]{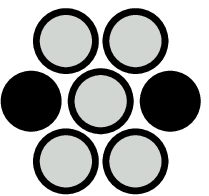}    & -2.4& -0.0&  0.7&  5.8&  6.3&  3.0&  1.3&  3.6&  2.5  \\
\includegraphics[height=10pt]{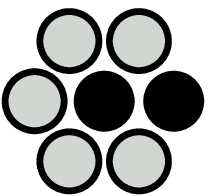}    & 57.0& 57.2& 47.5& 38.4& 43.7& 34.8&  1.4&  9.1&  6.6  \\
ml                                                  & 34.9& 48.9& 42.6& 33.4& 35.7& 26.0&  3.7&  7.7&  3.4  \\
%bulk                                                & \\ 
\end{tabular}
\end{ruledtabular}
\end{center}
\end{table}
%%%%%%%%%%%%%%%%%%%%%%%%%%%%%%%%%%%%%%%%%%%%%%%%%%%%%%%%%%%%%%%%%%%%%%%%%%%%%
The calculated isotropic exchange coupling constants $J_{ij}$  are presented
for all clusters (except for the cross-shaped 5-atom and 6-atom ones) in
Tab.~\ref{exchange_coupling}. Positive and negative values of $J_{ij}$
correspond to ferromagnetic and antiferromagnetic coupling, respectively.  In
Tab.~\ref{total_coupling} we show the sum of all couplings related to a
particular atom at site $i$.
%%%%%%%%%%%%%%%%%%%%%%%%%%%%%%%%%%%%%%%%%%%%%%%%%%%%%%%%%%%%%%%%%%%%%%%%%%%%%%%%
\begin{equation}
J^{i} =  \sum_{i \neq j} J_{ij} \;.
\end{equation}
%%%%%%%%%%%%%%%%%%%%%%%%%%%%%%%%%%%%%%%%%%%%%%%%%%%%%%%%%%%%%%%%%%%%%%%%%%%%%%%%
The  effective exchange field $J^{i}$ can be seen as the total strength by
which the magnetic moment at site $i$ is held along its direction by all other
atoms.   
%%%%%%%%%%%%%%%%%%%%%%%%%%%%%%%%%%%%%%%%%%%%%%%%%%%%%%%%%%%%%%%%%%%%%%%%%%%%%
\begin{table}%[b]
\caption{Effective exchange field $J^{i}$ in meV for Fe, Co and Ni clusters
         deposited on Ir(111), Pt(111) and Au(111). The icons in the left column
         indicate the corresponding cluster geometry as well as the cluster site
         $i$. The last line gives $J^{i}$ for the full monolayer (ml).  
          }
\label{total_coupling}
\begin{center}
\begin{ruledtabular}
\begin{tabular}{crrrrrrrrrr}
&\multicolumn{3}{c}{Fe}&\multicolumn{3}{c}{Co}&\multicolumn{3}{c}{Ni} \\
%\cmidrule(r){2-4} \cmidrule(r){5-7} \cmidrule(r){8-10}
$i$     &\multicolumn{1}{c}{Ir}&\multicolumn{1}{c}{Pt}&\multicolumn{1}{c}{Au}
        &\multicolumn{1}{c}{Ir}&\multicolumn{1}{c}{Pt}&\multicolumn{1}{c}{Au}
        &\multicolumn{1}{c}{Ir}&\multicolumn{1}{c}{Pt}&\multicolumn{1}{c}{Au} \\
%\midrule
\includegraphics[height=10pt]{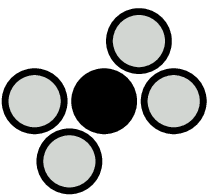}  &267.1&297.2&289.6&210.6&236.6&203.8& 16.4& 62.0& 42.1  \\
\includegraphics[height=10pt]{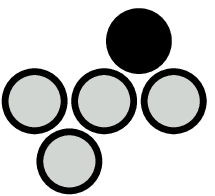}  &190.7&201.3&196.1&131.8&147.4&132.0&  6.9& 39.8& 30.6  \\
\includegraphics[height=10pt]{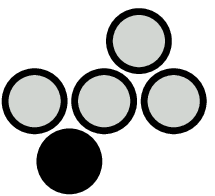}  &153.8&180.1&180.3&126.6&142.5&128.0&  9.1& 40.1& 30.2  \\
\includegraphics[height=10pt]{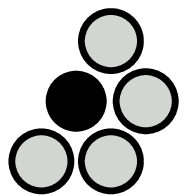}   &296.5&317.8&290.8&204.3&228.9&182.9&  9.6& 51.3& 37.7  \\
\includegraphics[height=10pt]{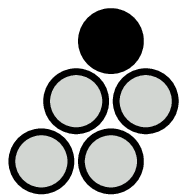}   &174.7&188.9&169.4&126.0&145.7&127.4&  8.7& 41.3& 25.9  \\
\includegraphics[height=10pt]{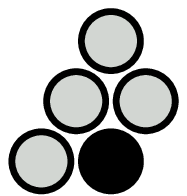}   &252.1&266.8&271.0&168.6&199.6&165.7&  6.7& 41.5& 27.6  \\
\includegraphics[height=10pt]{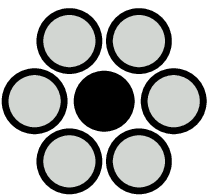}   &348.3&351.8&286.3&235.6&267.1&208.9&  8.7& 55.8& 39.9  \\
\includegraphics[height=10pt]{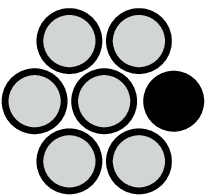}   &215.1&230.2&232.0&151.2&184.6&149.8& 11.3& 44.5& 26.9  \\
ml                                               &205.5&285.4&244.6&227.0&233.5&190.9& 27.8& 58.5& 27.7  \\
\end{tabular}
\end{ruledtabular}
\end{center}
\end{table}
%%%%%%%%%%%%%%%%%%%%%%%%%%%%%%%%%%%%%%%%%%%%%%%%%%%%%%%%%%%%%%%%%%%%%%%%%%%%%

From the data given in Tab.~\ref{exchange_coupling} one can see that Fe and Co
clusters show a strong ferromagnetic nearest neighbour coupling in the range of
40-140~meV while for Ni clusters the $J_{ij}$ values are often much smaller.
In the case of Fe and Co the couplings between nearest neighbouring atoms are
about one order of magnitude larger than couplings between more distant atoms,
i.e.\ the coupling strength falls off very rapidly with increasing the
interatomic distance.  For Ni clusters, however, and especially for Ni on
Ir(111) where the couplings are very weak this trend is less pronounced.  These
results also show that there is an occasional weak anti-ferromagnetic coupling
between more distant atoms for Fe clusters, which however, gives only an
insignificant contribution to the cummulative $J^{i}$ of each respective atom.
As each $J_{ij}$ contains by definition (see the Heisenberg Hamiltonian in
Eq.~(\ref{Hspin_2})) the product between the involved spin magnetic moments
$\mu^i_{\mathrm{spin}}$ and $\mu^j_{\mathrm{spin}}$ the coupling is, naturally,
largest for Fe and smallest for Ni clusters.  Moreover, the nearest neighbour
exchange coupling among the Fe, Co and Ni cluster atoms is larger than our
corresponding values for standard bcc Fe (37.8~meV), hcp Co (26.3~meV) and fcc
Ni (4.8~meV). 

Apart from the magnitude of the spin magnetic moments also atomic coordination
as well as substrate effects play an important role.  Especially for Fe
clusters the $J_{ij}$ values between low coordinated cluster atoms are often
much larger when compared to atoms with higher coordination.  Nevertheless, the
effective exchange field $J^{i}$ increases monotonically with increasing
coordination, i.e.\ given a fixed number of Fe, Co or Ni atoms the most compact
structure will form the most stable ferromagnetic state.  

As one can clearly see from the data in Table~\ref{exchange_coupling} in
combination with the cluster geometries given in
Figs.~\ref{spin_cluFe}-\ref{spin_cluNi} the isotropic exchange coupling is also
affected by the arrangement of cluster atoms with respect to the underlying
surface sites.  Looking at the two different compact Fe and Co trimers on
Ir(111) and Pt(111) for instance we find that the coupling values differ by
about 8-10\%.  For the 7-atom Fe cluster on Ir(111) and Pt(111), however,
$J_{ij}$ for nearest neighbouring edge atoms varies by as much as 20-45\%,
respectively, whereas the corresponding couplings for clusters on Au(111) do in
general not exhibit such a pronounced dependence on the atomic position with
respect to the substrate atoms.  The latter seems also to be the case for Fe
and Co 7-atom clusters with an identical configuration on Cu(111). This was
studied in detail by Mavropolus et al.~\cite{MLB10} and the Cu substrate atoms
also do not seem to participate in the exchange coupling of the Fe and Co
cluster atoms.  Therefore, we ascribe this substrate effect to the large
spin-polarisation within the the Ir and Pt surface atoms while the weak induced
magnetism  in Cu and Au causes only minor variations in the exchange coupling
of equidistant cluster atoms.  These irregularities in the couplings underline
that transferring $J_{ij}$ coupling constants obtained from bulk calculations
to low-dimensional finite nanostructures will lead in general to unreliable
results.

For Fe and Co clusters the magnitude of the isotropic exchange interaction is
quite similar for all three investigated substrates. Ni clusters, on the other
hand, have comparable nearest neighbour $J_{ij}$ values only when being
deposited on Pt(111) and Au(111) while deposition on Ir(111) reduces the
coupling strength to just a few meV.  This results in a quite small effective
exchange field $J^{i}$ per Ni atom  in the order of about 10~meV. As the
exchange interaction is so small for these Ni clusters, there is a pronounced
tendency that their magnetic ground state deviates strongly from a collinear
configuration (see below).

The coupling of magnetic cluster atoms to the induced magnetic moments in the
substrate is always very small. $J_{ij}$ is about 2~meV between Fe or Co
cluster atoms and topmost layer atoms of an Ir or Pt surface. The small induced
moments in the Au(111) substrate couple anti-ferromagnetically to the cluster
atoms. Here the nearest neighbour $J_{ij}$'s are only in the order of 0.1 meV
being of similar magnitude as the ferromagnetic coupling of Ni cluster atoms to
Ir or Pt surface sites.

In addition to the isotropic $J_{ij}$ coupling constants Tab.~\ref{dm} shows
the complementary data for the anisotropic exchange interaction. For clarity we
present here only the magnitude of the DM vector $|\vec{D}_{ij}|$ which can be
seen as a measure of the driving force towards a non-collinear spin
configuration. Given the fact that the SOC strength is comparable in Ir, Pt and
Au one can see from the data in Tab.~\ref{dm} that for any given cluster there
are often strong variations (without any clear trends) in $|\vec{D}_{ij}|$ upon
deposition onto different substrates. As discussed above for the $J_{ij}$
values, we find here an even more pronounced dependence of $|\vec{D}_{ij}|$ on
the position of cluster atoms with respect to the underlying substrate atoms
and the results show in addition that the relative decay of the DM interaction
with increasing interatomic distance is much weaker when compared to the
corresonding isotropic exchange coupling. 
%%%%%%%%%%%%%%%%%%%%%%%%%%%%%%%%%%%%%%%%%%%%%%%%%%%%%%%%%%%%%%%%%%%%%%%%%%%%%
\begin{table}%[b]
\caption{Anisotropic exchange coupling parameter $|\vec{D}_{ij}|$ in meV for
Fe, Co and Ni clusters deposited on Ir(111), Pt(111) and Au(111). The icons in
the left column indicate the corresponding cluster geometry as well as the
cluster sites $i$ and $j$, respectively. The last line gives $|\vec{D}_{ij}|$
for nearest neighbouring sites within the full monolayer (ml). }
\label{dm}
\begin{center}
\begin{ruledtabular}
\begin{tabular}{crrrrrrrrrr}

&\multicolumn{3}{c}{Fe}&\multicolumn{3}{c}{Co}&\multicolumn{3}{c}{Ni} \\
%\cmidrule(r){2-4} \cmidrule(r){5-7} \cmidrule(r){8-10}
$ij$    &\multicolumn{1}{c}{Ir}&\multicolumn{1}{c}{Pt}&\multicolumn{1}{c}{Au}
        &\multicolumn{1}{c}{Ir}&\multicolumn{1}{c}{Pt}&\multicolumn{1}{c}{Au}
        &\multicolumn{1}{c}{Ir}&\multicolumn{1}{c}{Pt}&\multicolumn{1}{c}{Au} \\
%\midrule
\includegraphics[height=10pt]{2atom1_2.eps}    &1.17 & 6.93 & 1.61 & 3.48 & 5.47 & 2.34 & 0.26 & 0.24 & 0.49  \\
\includegraphics[height=10pt]{3c2atom1_2.eps}  &4.60 & 6.16 & 0.70 & 1.06 & 1.17 & 4.18 & 0.39 & 1.45 & 0.99  \\
\includegraphics[height=10pt]{3c1atom1_2.eps}  &0.94 & 6.31 & 2.45 & 5.76 & 8.31 & 8.66 & 0.24 & 2.35 & 1.32  \\
\includegraphics[height=10pt]{3latom1_2.eps}   &1.83 & 5.64 & 2.77 & 2.06 & 3.51 & 1.27 & 0.48 & 0.63 & 0.48  \\
\includegraphics[height=10pt]{3latom2_3.eps}   &3.62 & 1.19 & 3.78 & 0.38 & 1.79 & 2.91 & 0.40 & 1.20 & 1.42  \\
\includegraphics[height=10pt]{4atom1_2.eps}    &4.86 & 6.02 & 1.52 & 2.29 & 0.67 & 1.73 & 0.14 & 0.97 & 1.45  \\
\includegraphics[height=10pt]{4atom1_3.eps}    &1.76 & 5.64 & 2.26 & 4.33 & 4.24 & 2.85 & 0.20 & 1.09 & 1.57  \\
\includegraphics[height=10pt]{4atom1_6.eps}    &2.54 & 5.30 & 0.98 & 1.19 & 1.39 & 2.14 & 0.26 & 0.28 & 0.39  \\
\includegraphics[height=10pt]{4atom3_6.eps}    &1.58 & 0.64 & 0.75 & 0.34 & 1.75 & 0.99 & 0.09 & 0.58 & 0.19  \\
\includegraphics[height=10pt]{5atom_d_1_6.eps} &5.75 & 5.81 & 0.63 & 2.97 & 6.03 & 7.98 & 0.54 & 0.87 & 0.26  \\
\includegraphics[height=10pt]{5atom_d_1_7.eps} &2.25 & 5.12 & 1.85 & 4.81 & 5.63 & 3.29 & 0.52 & 1.42 & 0.73  \\
\includegraphics[height=10pt]{5atom_d_2_6.eps} &4.67 & 6.97 & 2.51 & 3.43 & 6.24 & 5.82 & 0.51 & 0.88 & 0.90  \\
\includegraphics[height=10pt]{5atom_d_5_7.eps} &1.39 & 5.46 & 1.45 & 7.55 & 9.98 & 4.72 & 0.50 & 1.06 & 1.48  \\
\includegraphics[height=10pt]{5atom_d_2_7.eps} &1.30 & 1.03 & 2.24 & 0.83 & 0.80 & 1.95 & 0.37 & 0.63 & 0.23  \\
\includegraphics[height=10pt]{5atom_d_6_7.eps} &2.34 & 0.77 & 0.86 & 1.40 & 2.21 & 3.11 & 0.16 & 0.61 & 0.40  \\
\includegraphics[height=10pt]{5atom1_2.eps}    &5.60 & 5.26 & 2.00 & 2.30 & 1.82 & 5.39 & 0.15 & 0.88 & 1.21  \\
\includegraphics[height=10pt]{5atom1_4.eps}    &3.33 & 3.64 & 1.19 & 1.14 & 1.47 & 1.34 & 0.26 & 0.54 & 0.34  \\
\includegraphics[height=10pt]{5atom2_3.eps}    &2.51 & 5.22 & 1.43 & 3.83 & 1.54 & 3.81 & 0.39 & 0.42 & 1.19  \\
\includegraphics[height=10pt]{5atom2_7.eps}    &2.90 & 4.90 & 1.21 & 1.46 & 2.58 & 2.40 & 0.31 & 0.28 & 0.76  \\
\includegraphics[height=10pt]{5atom2_4.eps}    &0.79 & 0.35 & 0.73 & 0.40 & 1.73 & 1.10 & 0.17 & 0.40 & 0.49  \\
\includegraphics[height=10pt]{5atom7_4.eps}    &1.49 & 0.46 & 1.45 & 1.12 & 1.31 & 1.74 & 0.35 & 0.74 & 0.27  \\
\includegraphics[height=10pt]{7atom2_3.eps}    &1.00 & 4.99 & 1.08 & 5.01 & 1.86 & 1.99 & 0.45 & 0.26 & 0.72  \\
\includegraphics[height=10pt]{7atom2_7.eps}    &3.72 & 6.00 & 2.55 & 1.44 & 2.88 & 1.56 & 0.26 & 0.23 & 1.22  \\
\includegraphics[height=10pt]{7atom2_4.eps}    &1.28 & 0.43 & 0.97 & 0.57 & 1.20 & 0.80 & 0.10 & 0.27 & 0.39  \\
\includegraphics[height=10pt]{7atom2_6.eps}    &1.28 & 0.43 & 0.97 & 0.57 & 1.20 & 0.84 & 0.10 & 0.29 & 0.38  \\
\includegraphics[height=10pt]{7atom2_5.eps}    &0.58 & 0.43 & 1.70 & 1.97 & 1.56 & 0.86 & 0.12 & 0.33 & 0.33  \\
\includegraphics[height=10pt]{7atom1_2.eps}    &3.81 & 2.52 & 2.55 & 1.48 & 1.75 & 1.67 & 0.13 & 0.35 & 0.45  \\
ml                                                  &4.17 & 2.54 & 0.66 & 1.33 & 1.68 & 0.71 & 0.13 & 0.20 & 0.04  \\
%bulk                                                & \\ 
\end{tabular}
\end{ruledtabular}
\end{center}
\end{table}
%%%%%%%%%%%%%%%%%%%%%%%%%%%%%%%%%%%%%%%%%%%%%%%%%%%%%%%%%%%%%%%%%%%%%%%%%%%%%

Albeit that $|\vec{D}_{ij}|$ is between one or two orders of magnitude smaller
than the isotropic exchange coupling, it is not negligible. For Fe$_n$ on
Ir(111) we obtain a relatively strong DM interaction which is in accordance
with the recent findings of Heinze et al.~\cite{MLB+10} and of von Bergmann et
al.~\cite{BHB+06} as well as De\'ak et al.~\cite{DSU11} for Fe/Ir(001) which
all demonstrate that these systems show a strong tendency towards non-collinear
magnetism.  Moreover, our results also show large $|\vec{D}_{ij}|$ values for
Fe$_n$ and Co$_n$ clusters from which we conjecture that this may also lead to
complex magnetic structures within extended Fe and Co nanostructures on these
substrates. In fact the sometimes experimentally observed, unexpected low
magnetic moments in Fe- and Co-Pt(111) systems may be caused by this
mechanism~\cite{HLK+09,SKZ+10}.  For Ni clusters the DM interaction is always
very important with respect to the isotropic exchange coupling as both
quantities are often of the same order of magnitude.  Thus, one can expect the
presence of  non-collinear magnetic ordering in Ni clusters on all three
substrates. It should be stressed that this non-collinearity will {\em not} be
a consequence of the frustration between the magnetic and geometric order but
rather will follow from the influence of spin-orbit effects on the exchange
coupling, as manifested by the DM interaction.

\subsection{Magnetic Anisotropy Energy}

%%%%%%%%%%%%%%%%%%%%%%%%%%%%%%%%%%%%%%%%%%%%%%%%%%%%%%%%%%%%%%%%%%%%%%%%%%%%%
\begin{table}%[t,b]
\caption{Magnetic anisotropy energy (MAE) per atom for Fe, Co and Ni clusters
deposited on Ir(111), Pt(111) and Au(111). The icons in the left column
indicate the corresponding cluster geometry. The positive (negative) values of
the MAE (in meV) correspond to an out-of-plane (in-plane) magnetic easy axis.
For the full monolayers the total MAE $\Delta E$ is decomposed into its dipolar
part $\Delta E_{\text{dd}}$ and its magnetocrystalline part $\Delta
E_{\text{soc}}$.  The latter is further decomposed into contributions that
originate from the monolayer ($\Delta E_{\text{soc}}^{3d}$) and the substrate
($\Delta E_{\text{soc}}^{5d}$), respectively. For the deposited clusters we
found $\Delta E_{\text{dd}}\approx0$ and $\Delta E_{\text{soc}}\approx \Delta
E_{\text{soc}}^{3d}$.}
\label{tab:clu_MAE}
\begin{center}
\begin{ruledtabular}
\begin{tabular}{crrrrrrrrrr}

&\multicolumn{3}{c}{Fe}&\multicolumn{3}{c}{Co}&\multicolumn{3}{c}{Ni} \\
%\cmidrule(r){2-4} \cmidrule(r){5-7} \cmidrule(r){8-10}
        &\multicolumn{1}{c}{Ir}&\multicolumn{1}{c}{Pt}&\multicolumn{1}{c}{Au}
        &\multicolumn{1}{c}{Ir}&\multicolumn{1}{c}{Pt}&\multicolumn{1}{c}{Au}
        &\multicolumn{1}{c}{Ir}&\multicolumn{1}{c}{Pt}&\multicolumn{1}{c}{Au} \\
%\midrule
\includegraphics[height=10pt]{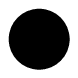}   & 0.10& 8.42&11.45& 3.95& 4.88& 9.02&-0.21&-1.57&-5.11  \\
\includegraphics[height=10pt]{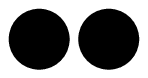}   &-1.01& 0.48& 2.75& 0.54& 2.25&-0.39& 0.10& 0.08&-1.17  \\
\includegraphics[height=10pt]{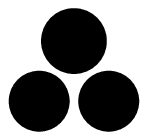} & 0.37& 0.36& 1.33& 0.07&-0.12&-2.45& 0.84& 0.49& 0.24  \\
\includegraphics[height=10pt]{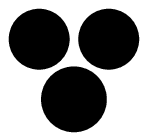} &-0.45& 0.33& 1.44& 1.82& 2.00& 0.52& 0.17&-0.13&-0.62  \\
\includegraphics[height=10pt]{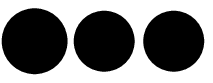} &-0.07& 1.23& 3.46&-0.23& 0.60&-3.84& 0.10&-0.58& 3.08  \\
\includegraphics[height=10pt]{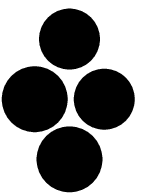}   &-0.09& 0.36& 1.74& 0.04&-0.18&-2.56& 0.09& 0.44& 0.15  \\
\includegraphics[height=10pt]{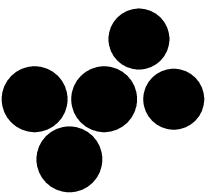} &-0.49& 0.05& 1.26& 0.74& 1.42&-0.03& 0.14& 0.23&-0.77  \\
\includegraphics[height=10pt]{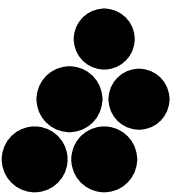} & 0.11& 0.49& 1.65& 0.17&-0.44&-2.32& 0.05& 0.21& 0.21  \\
\includegraphics[height=10pt]{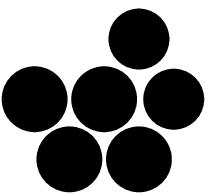}   & 0.09& 0.48& 1.90& 0.36&-0.11&-2.00& 0.08& 0.22& 0.09  \\
\includegraphics[height=10pt]{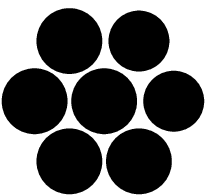}   & 0.90& 2.15& 3.86& 0.30&-0.26&-2.12& 0.11& 0.37& 0.65  \\
                                                & \multicolumn{9}{c}{monolayer}                         \\
$\Delta E$                                      & 0.64&-0.26& 1.09& 0.20& 0.31&-1.32&-0.19&-0.39&-0.44  \\
$\Delta E_{\text{dd}}$                          &-0.19&-0.19&-0.18&-0.09&-0.09&-0.08&-0.01&-0.01&-0.01  \\
$\Delta E_{\text{soc}}$                         & 0.83&-0.07& 1.27& 0.29& 0.40&-1.24&-0.18&-0.38&-0.43  \\
$\Delta E_{\text{soc}}^{3d}$                    & 0.82&-0.17& 1.27& 0.30& 0.22&-1.24&-0.19&-0.46&-0.43  \\
$\Delta E_{\text{soc}}^{5d}$                    & 0.01& 0.10& 0.00&-0.01& 0.18& 0.00& 0.01& 0.08& 0.00  \\
%%bulk  & \\ 
\end{tabular}
\end{ruledtabular}
\end{center}
\end{table}
%%%%%%%%%%%%%%%%%%%%%%%%%%%%%%%%%%%%%%%%%%%%%%%%%%%%%%%%%%%%%%%%%%%%%%%%%%%%%
The magnetic anisotropy energies (MAE) per atom for all investigated clusters
are compiled in Table~\ref{tab:clu_MAE}.  Positive MAE values denote an
out-of-plane anisotropy while negative MAE values correspond to an in-plane
magnetic easy axis. Fe clusters on Pt(111) and Au(111) show always an
out-of-plane MAE whereas all other cluster substrate systems exhibit a rather
nonuniform behaviour of their MAE with varying cluster size or geometry.  This
complex behaviour arises from the fact that already tiny changes in the
electronic structure can cause large changes in the MAE. This can be seen again
for example in the case of the compact trimers where one can observe a dramatic
dependence of the MAE on their position with respect to the substrate, i.e.
depending on whether a substrate atom is underneath the cluster centre or not.

All dimers and linear trimers with in-plane MAE have their magnetic easy axis
fixed along the cluster axis which is a result of the strong azimuthal MAE in
these systems being in the order of 1-4~meV per atom. For compact symmetric
clusters as well as for the full monolayers there remains only a very small
azimuthal MAE in the order of~$\mu$eV being thus negligible.

When evaluating the MAE by means of the torque method, contributions stemming
from all individual atomic sites of the system are added together.  One can
therefore technically identify which portion of the MAE comes from the adsorbed
atoms and which portion comes from the substrate atoms.  We found that the
contribution coming from the substrate is negligible in the case of clusters
while it can be significant  in the case of complete monolayers (e.g. up to
45\% of the total value for the Co monolayer on the Pt(111) substrate). This is
plausible given the fact that for monolayers, the substrate atoms are subject
to interaction with a larger number of adsorbed atoms, meaning that their
spin-polarization will be stronger and more robust than in the case of small
clusters, contributing thereby more significantly to the MAE.  At the same
time, one has to bear in mind that energy is not an extensive quantity and that
any decomposion of the MAE into parts has only a limited significance.

Concerning the dipole-dipole or shape MAE contribution, for clusters it is
negligible while for complete monolayers it attains appreciable values of
\mbox{-0.19}~meV and \mbox{-0.09}~meV per atom for Fe and Co monolayers,
respectively.  Moreover, we find that the substrate as well as the
dipole-dipole contribution to the MAE is negligible for clusters, whereas for
monolayers both contribution are much more important.

\subsection{Comparison with other works}

As already mentioned in the introduction it is not always straightforward to
directly compare theoretical LSDA results obtained by different computational
{\em ab-initio} implementations due to differences in the truncation of the
wavefunction or Green's function etc. as well as different technical issues and
approches as for example the implementation of spin-orbit coupling as
perturbation, the use of a supercell vs. embedding techniques or approximations
in the description of the effective potentials and so forth.  All this can
affect the obtained numerical results, especially  for sensitive magnetic
quantities like for instance orbital magnetic moments and magnetic anisotropy
energies.

Among the cluster/substrate systems discussed in this article, only Fe$_1$ and
Co$_n$ on Pt(111) have been studied extensively by other groups and we find for
these systems that our spin magnetic moments agree quantitatively well with the
corresponding $\mu_{\rm spin}$ values given in
Refs.~\cite{GRV+03,EZWV08,LSW03,BST+09,SCL+03,CFB08,BH09} using identical
geometries. The same is true for Fe$_3$ on Pt(111)~\cite{RGDP11}, Fe$_1$ and
Co$_1$ on Ir(111)~\cite{EZWV08} as well as for the monolayer systems
Fe/Ir(111)~\cite{EZWV08,BHB+06}, Co/Ir(111)~\cite{EZWV08}, Fe/Pt(111),
Co/Pt(111)~\cite{EZWV08} and Co/Au(111)~\cite{USBW96}.  Regarding the values of
$\mu_{\rm orb}$ and the MAE, however, the agreement is in general less good,
i.e.\ only qualitative or worse, for the above mentioned reasons. As already
analysed by ~\v{S}ipr et al.~\cite{SBME11}, methods which rely on a supercell
approach~\cite{SCL+03,CFB08,BH09} produce always significantly higher induced
spin magnetic moments within the substrate atoms when compared to methods which
apply embedding techniques~\cite{GRV+03,EZWV08,LSW03,BST+09}.

\section{Summary and Conclusions}

The evolution of the spin and orbital magnetic moments of the investigated 3$d$
transition metal clusters on 5$d$ noble metal surfaces mostly follows common
trends and patterns that can be understood by considering the coordination
numbers of atoms in the clusters and the polarizability of the substrate.  The
average $\mu_{\rm spin}$ values decrease nearly monotonously with the number of
atoms in the cluster being at variance with trends observed for free
clusters~\cite{SME09a}.  Our results show that $\mu_{\rm orb}$ may strongly
depend on the position of the cluster with respect to the surface atoms, as
demonstrated in particular for the triangular 3-atom clusters on Pt(111) and
Au(111).  The magnetic moments for Ni clusters on Ir are smaller than one would
expect judging from the trends for the other cluster/substrate combinations.
Moreover, they depend wildly on number of atoms in the cluster and their
smallness is compatible with the fact that the peak in minority DOS is below
E$_{\rm F}$.

Apart from Ni$_n$/Ir(111) all clusters show a strong ferromagnetic isotropic
exchange coupling exceeding the corresponding bulk values of standard bcc Fe,
hcp Co and fcc Ni. In addition, there are also strong anisotropic DM
interactions present revealing the intrinsic tendencies towards noncollinear
magnetism in these systems.  Finally, the magnetic anisotropy energies can be
very large for some cluster/substrate or surface/substrate combinations, but
unfortunately, there are no clear trends visible that would allow any
straightforward anticipation of this sensitive quantity.

\acknowledgments
Financial support by the Bundesministerium f\"ur Bildung und Forschung (BMBF)
Verbundprojekt R\"ont\-gen\-ab\-sorptions\-spektroskopie (05K10WMA und 05K10GU5), 
Deutsche Forschungsgemeinschaft (DFG) via SFB668 and by the Grant
Agency of the Czech Republic (108/11/0853) is gratefully acknowledged.

\bibliographystyle{prsty}

%\bibliography{thesis}

\end{document}